\documentclass[a4paper,12pt]{article}
\usepackage[T1]{fontenc}
\usepackage{amsmath}
\usepackage{amssymb}
\usepackage{cite}
\usepackage{graphicx}
\usepackage{a4wide}
\usepackage{float}
\usepackage[small]{caption}
\DeclareMathOperator{\tr}{tr}
\newcommand{\E}[1]{\ensuremath{\mathrm{E}_{#1}}} 
\newcommand{\SU}[1]{\ensuremath{\mathrm{SU}(#1)}}
\newcommand{\SO}[1]{\ensuremath{\mathrm{SO}(#1)}}
\newcommand{\U}[1]{\ensuremath{\mathrm{U}(#1)}}
\newcommand{\Z}[1]{\ensuremath{\mathbb{Z}_{#1}}} 

\begin{document}

\date{}
\title{
\begin{flushright}
\normalsize{DESY 10-111}\\
\normalsize{LMU-ASC 55/10}\\
\end{flushright}
\vskip 1.6cm
{\bf\LARGE Proton Hexality in Local Grand Unification}\\[0.8cm]}

\author{{\bf\normalsize
Stefan~F\"orste$^a$\!,
Hans~Peter~Nilles$^a$\!,
Sa\'ul~Ramos-S\'anchez$^b$} 
\\{\bf\normalsize and Patrick~K.S.~Vaudrevange$^c$}\\[1cm]
{\it\normalsize
${}^a$ Bethe Center for Theoretical Physics and}\\[-0.5mm]
{\it\normalsize Physikalisches Institut der Universit\"at Bonn,}\\[-0.5mm]
{\it\normalsize Nussallee 12, 53115 Bonn, Germany}\\[2mm]
{\it\normalsize
${}^b$ Deutsches Elektronen-Synchrotron DESY, Hamburg, Germany}\\[2mm]
{\it\normalsize
${}^c$ Arnold Sommerfeld Center for Theoretical Physics,}\\[-0.05cm]
{\it\normalsize Ludwig-Maximilians-Universit\"at M\"unchen, 80333
  M\"unchen, Germany}
}

\maketitle

\thispagestyle{empty}

\vspace{1cm}

\begin{abstract}
\noindent

Proton hexality is a discrete symmetry that avoids the problem of
too fast proton decay in the supersymmetric extension of the
standard model. Unfortunately it is inconsistent with 
conventional grand unification. We show that proton hexality
can be incorporated in the scheme of  ``Local Grand Unification''
discussed in the framework of model building in
(heterotic) string theory.

\end{abstract}

\newpage
\section{Introduction}

The question of proton stability is of great importance in elementary
particle physics. Conserved baryon number (B) could be a reason for a
stable proton but would be incompatible with the creation of a baryon
asymmetry in the universe. In the standard $\SU3\times\SU2\times\U1$
model of particle physics $\U1_B$ is a good symmetry at the
renormalisable level (broken by non-perturbative effects and possibly
dimension-6 operators) and proton decay is sufficiently suppressed.
The phenomenon of B-violation is thus mainly concerned with physics beyond the
standard model (SM).  Indeed, a natural framework to address the question
is grand unification (GUTs). Quarks and leptons appear in
unified multiplets and $\U1_B$ is broken. The known stability of the
proton requires the GUT scale to be rather large $M_{GUT} > 10^{16}$
GeV such that dimension-6 operators are sufficiently suppressed.

Over the years it appears that the framework of GUTs seems to require
supersymmetry. This is then consistent with unification of gauge coupling
constants at $M_{GUT} \sim 10^{16}$ but it leads to new complications
with proton stability due to potentially dangerous dimension-4 and
dimension-5 operators. New symmetries like R-parity~\cite{Farrar:1978xj}
(or matter parity~\cite{Dimopoulos:1981zb,Dimopoulos:1981dw})
have been conjectured to forbid the dim-4
operators.  Such a symmetry would lead to a new stable particle as the
source for dark matter.  Still, the dim-5 operators are problematic.
Other discrete symmetries like baryon triality
($B_3$)~\cite{Ibanez:1991hv,Ibanez:1991pr} might solve the problem.
The most attractive symmetry is proton hexality ($P_6$) and has been
identified by Dreiner, Luhn and Thormeier~\cite{Dreiner:2005rd}. 
It forbids all the problematic
dim-4 and -5 operators, but allows lepton number violation in form of
Majorana neutrino masses: thus $P_6$ perfectly fits our
needs.

Proton hexality is a beautiful symmetry and grand unification is a
very attractive scheme: but unfortunately there is a clash between the
two. $P_6$ is incompatible with a unified structure of quark and
lepton multiplets~\cite{Luhn:2007gq}. This is in contrast to matter parity $\Z2^\textrm{matter}$
which can be incorporated e.g. in an \SO{10} GUT (it is a discrete
subgroup of $\U1_{B-L}$ in \SO{10}).  This is not true for $B_3$ and
$P_6$. An ultraviolet (UV) completion of theories with $P_6$ needs
something more general than grand unification. String theory could be
a candidate as it also includes a consistent description of
gravitational interactions.\footnote{Global symmetries might be broken
by gravitational interactions. Therefore it is important to discuss
these questions in theories where gravity is consistently incorporated.}

Recent work towards string theoretic constructions of the minimal
supersymmetric standard model (MSSM)~\cite{Buchmuller:2005jr,Buchmuller:2006ik,Lebedev:2006kn,Lebedev:2007hv,Lebedev:2008un} has revealed the concept
of ``Local Grand Unification''~\cite{Kobayashi:2004ya,Buchmuller:2005jr,Forste:2004ie}, a variant of GUTs that addresses
some of its problematic properties. It allows ``split multiplets'' that
e.g. solve the doublet-triplet splitting in the Higgs-sector and
simplifies the breakdown of the grand unified gauge groups. These
incomplete or split multiplets make it possible that $P_6$ can become
compatible with local grand unification.

Moreover, it has recently been argued that discrete symmetries appear
abundantly in string model constructions~\cite{Kobayashi:2006wq,Araki:2008ek,Petersen:2009ip}, 
with important applications for
particle physics model building~\cite{Ko:2007dz,Buchmuller:2008uq,Lebedev:2009ag,Ishimori:2010au,Kappl:2010xx}.
We are thus in a situation that
$P_6$ could originate from a string model as consistent UV
completion and that such a symmetry is compatible with local grand
unification. Within this top-down approach we can be confident that
the global symmetries are respected by gravitational 
interactions.\footnote{Earlier attempts in a bottom-up 
    approach~\cite{Ibanez:1991hv,Ibanez:1991pr,Dreiner:2005rd,Luhn:2007gq} tried to
    solve this problem through the notion of ``anomaly free discrete
    symmetries''. In the top-down approach we do not have to worry
    about these constraints as long as we are dealing with a
    consistent string model.}

The present paper is devoted to the study of proton hexality within
the framework of local grand unification. 
In section 2 we shall
present $P_6$ followed in section 3 by a discussion of the
incompatibility of $P_6$ with GUTs. 
In section 4 we shall try to
incorporate $P_6$ via a bottom-up approach in extra-dimensional
GUT-like theories and stress the geometric aspects of local grand
unification. Section 5 presents the results of heterotic orbifold
constructions towards the incorporation of $P_6$. We provide some toy
models where $P_6$ appears in various different ways, anomalous or
non-anomalous. A completely satisfactory model has not been found yet.
Section 6 will discuss possible lines of future research in the
direction of explicit model building.

\section{Proton Hexality}

In this section we will motivate proton hexality as a discrete
symmetry in the supersymmetric extension of the SM in
somewhat more detail. For derivations, however, we will refer to
original literature. 

Ensuring sufficient stability of the
proton in theories beyond the SM often provides non-trivial 
restrictions. New fields can give rise to baryon or lepton 
number violating couplings. In supersymmetric extensions
of the SM, for instance, the most general superpotential
respecting renormalisability and gauge invariance is
\begin{align}
W =\quad & h^E _{ij}\, L_i H_d \overline{E}_j + h^D_{ij}\, Q_i H_d
\overline{D}_j + h_{ij}^U\, Q_i H_u \overline{U}_j + \mu\, H_d H_u
\nonumber\\
& + \lambda_{ijk}\, L_i L_j \overline{E}_k +\lambda^\prime _{ijk}\, L_i
Q_j \overline{D}_k + \kappa_i\, L_i H_u \nonumber \\
& + \lambda^{\prime\prime}_{ijk}\, \overline{U}_i
\overline{D}_j\overline{D}_k , \label{eq:suppotren}
\end{align}
where $i,j,k$ are family indices and gauge indices are suppressed. 
Terms in the first line encode Yukawa couplings needed for lepton and
quark mass generation and the $\mu$-term contribution to the Higgs
potential. Terms in the second and third line of~\eqref{eq:suppotren}
violate lepton and baryon number, respectively. 
The lepton or baryon number violating couplings ($\lambda$,
$\lambda^\prime$, $\kappa$ and $\lambda^{\prime\prime}$) 
can be forbidden by imposing an additional discrete
symmetry such as $R$-parity or matter parity. Compared to $R$-parity, matter parity is a
\Z2 symmetry under which all constituents of a chiral
multiplet carry the same charge, {\it viz.}\ matter multiplets are odd
whereas Higgs multiplets are even.  

Assuming that the supersymmetric extension of the SM
originates from a more fundamental theory, such as a GUT, we also have
to discuss effective non-renormalisable couplings. 
The superpotential can contain dangerous terms
respecting all the symmetries including matter parity, 
\begin{equation}
\kappa^{(1)}_{ijkl}\, Q_i Q_j Q_k L_l + \kappa^{(2)}_{ijkl}\,
\overline{U}_i\overline{U}_j\overline{D}_k \overline{E}_l .
\end{equation}
These terms lead to dimension five  interactions violating baryon as
well as lepton number. 
Such dimension five couplings are suppressed by just
one power of the GUT scale and lead to too fast proton decay
\cite{Hinchliffe:1992ad}. The authors of~\cite{Ibanez:1991hv,Ibanez:1991pr}
proposed another discrete \Z3 symmetry forbidding also
the dimension five baryon number violating couplings. This symmetry
was later dubbed baryon triality, $B_3$ \cite{Martin:1997ns,Grossman:1998py}.
The charges are defined modulo hypercharge ($Y$) transformation and
listed in Table \ref{tab:Charges}. 

Matter parity as well as baryon triality 
can be obtained from a spontaneously broken additional \U1
gauge symmetry 
\cite{Ibanez:1991hv,Ibanez:1991pr}. Anomaly cancellation restricts the
spectrum of a gauge theory leading to a finite number of
possible discrete symmetries, which can be obtained in such a way. The
authors of \cite{Dreiner:2005rd} looked through all such symmetries
and identified proton hexality, $P_6$, as the only other
phenomenologically interesting discrete symmetry. On SM
fields $P_6$, as defined by the charge assignments in Table~\ref{tab:Charges},
acts as a \Z6 symmetry which is the product
of baryon triality and matter parity. 
Proton hexality had been discussed before in
\cite{Hamaguchi:1998wm,Babu:2002tx,Babu:2003qh,Wang:2004mg}. 
It forbids all dangerous dimension four and five baryon or lepton
number violating couplings while phenomenologically desirable
couplings are allowed. That is, only terms in the first line of~\eqref{eq:suppotren}
are allowed at the renormalisable level. At dimension five level, only interactions
coming from a superpotential $L H_u LH_u$ (family indices are
suppressed) are allowed. These respect baryon number
and, moreover, provide Majorana mass terms for left-handed neutrinos
after electroweak symmetry breaking. Dimension six
interactions are suppressed by two powers of the GUT scale which
results in a sufficiently stable proton for supersymmetric theories
(with $M_{GUT} \sim 10^{16}$ GeV).
\begin{table}[h!]
  \centering
  \begin{tabular}{|l||c|c|c|c|c|c|c|c|}
    \hline
   \phantom{$A^{A^A}$} & $Q$ & $\bar{U}$ & $\bar{D}$ & $L$ & $\bar{E}$ & $H_u$ & $H_d$ & $\bar{\nu}$ \\
    \hline\hline
    $6\ Y$& $1$ & $-4$ & $2$ & $-3$ & $6$ &  $3$ & $-3$ & $0$ \\
    \hline\hline     
    $\Z2^\textrm{matter}$ 
          & $1$ & $1$  & $1$ &  $1$ & $1$ &  $0$ &  $0$ & $1$ \\
    $B_3$ & $0$ & $-1$ & $1$ & $-1$ & $2$ &  $1$ & $-1$ & $0$ \\
    $P_6$ & $0$ & $1$  &$-1$ & $-2$ & $1$ & $-1$ &  $1$ & $3$ \\
    \hline
  \end{tabular}
  \caption{Hypercharge and discrete charges of the MSSM particles
    under hypercharge $Y$, matter parity 
    $\Z2^\textrm{matter}$, baryon triality  $B_3$ and proton
    hexality $P_6$. One 
 can show that
    $P_6=\Z2^\textrm{matter}\times B_3$ up to a hypercharge
    shift. $\Z2^\textrm{matter}$, $B_3$ and
    $P_6$ are defined just modulo $2$, $3$  and $6$, 
    respectively. A right-handed neutrino $\bar{\nu}$ has been included.}  
  \label{tab:Charges}
\end{table}
 
So far, our discussion did not include right-handed neutrinos. They 
can be included in a straightforward way (as shown in Table 1).
Interactions including right-handed neutrinos do not introduce baryon
number violating terms, since their $B_3$ charge is zero. All terms
needed for the see-saw mechanism are allowed by $P_6$~\cite{Dreiner:2005rd}. 

\section{Proton Hexality and Unified Gauge Groups}

Since proton hexality forbids also dangerous dimension five couplings,
it is desirable to embed $P_6$ into an underlying more fundamental
theory. As a first example we look at grand unification. The embedding
of $P_6$ into a GUT has been examined to some extent in~\cite{Luhn:2007gq}.  
There, the authors added an extra anomaly free
\U1 to the unified gauge group and identified all possible discrete
subgroups of that \U1. In this case, proton hexality does
not work with any of the usual candidate unified gauge groups
(Pati-Salam, \SU5, \SO{10}).

\subsection{Proton Hexality from Pati-Salam$\boldsymbol{\times\text{U}(1)_X}$}

An option which has been excluded in \cite{Luhn:2007gq} is to break the unified 
gauge group times an extra $\U1_X$ simultaneously to the SM gauge
group times $P_6$. For unified groups with rank larger than four $P_6$
can, in principle, be embedded diagonally into the extra $\U1_X$ times
another \U1 originating from the GUT gauge group. Such a scheme
works for Pati--Salam\footnote{Other discrete symmetries
  suppressing proton decay within Pati--Salam and \SO{10} have been
  identified in \cite{Mohapatra:2007vd}.}
with gauge group $\SU4\times\SU2_L\times\SU2_R$. The 
colour \SU3 is embedded into the upper three times
three block of the hermitian \SU4 generator and hypercharge
transformations are generated by a combination of an \SU4 and $\SU2_R$
transformation
\begin{equation}\label{eq:PSemb}
\text{SU}(3)\! :\, \left(\! \begin{array}{c c c  c}
\multicolumn{3}{c}{trace} &\! 0\!\\[-0.5ex]
\multicolumn{3}{c}{less} &\! 0\! \\[-0.5ex]
\multicolumn{3}{c}{} &\! 0\! \\[-0.5ex]
\!0 &\! 0\! &\! 0 &\! 0\!
\end{array}\!\right) ,\,\,\, \,\,\, \text{U}(1)_Y \! :\,
\left(\! \begin{array}{cccc} 
\!\frac{1}{6} &\! 0 &\! 0 &\! 0\!\\[-0.3ex]
\! 0 &\! \frac{1}{6}&\! 0 &\! 0\!\\[-0.3ex]
\! 0 &\! 0 &\! \frac{1}{6} &\! 0\!\\[-0.3ex]
\! 0 &\! 0 &\! 0 & \!\!\! -\frac{1}{2}\! \end{array}\!\right)
+\frac{1}{2}\left( \begin{array}{cc}\! 1 & \! 0\! \\
\! 0 &\!\!\! -1\!\end{array}\right)_R . 
\end{equation}
The left-handed quarks and leptons merge into a 
$(\boldsymbol{4},\boldsymbol{2},\boldsymbol{1})$ while the right-handed 
quarks and leptons form a 
$(\overline{\boldsymbol{4}},\boldsymbol{1},\boldsymbol{2})$ representation. 
The supersymmetric SM Higgs pair is combined into a 
$(\boldsymbol{1},\boldsymbol{2},\boldsymbol{2})$ representation. 
Now, we add an additional $\U1_X$ with generator $X$ and denote 
the corresponding charges by subscripts. With the assignments 
\begin{equation}\label{eq:assign}
(\boldsymbol{4},\boldsymbol{2},\boldsymbol{1})_{1}, \quad (\overline{\boldsymbol{4}},\boldsymbol{1},\boldsymbol{2})_{-1} \quad\text{and}\quad (\boldsymbol{1},\boldsymbol{2},\boldsymbol{2})_0  
\end{equation}
we reproduce the correct $P_6$ charges if we take $P_6$ to be the
${\mathbb Z}_6$ subgroup of the \U1 generated by
\begin{equation}\label{eqn:P6inPS}
P_6\! : \; \frac{1}{2}\left( \begin{array}{cccc}
\! 1 & \! 0 & \! 0 &\! 0 \! \\[-0.5ex]
\! 0 & \! 1 & \! 0 & \! 0 \!\\[-0.5ex]
\! 0 & \! 0 & \! 1 & \! 0 \! \\[-0.5ex]
\! 0 & \! 0 & \! 0 & \! -3\! \end{array}\right) -
\left( \begin{array}{cc}\! 1 & \! 0\! \\ 
\! 0 &\! -1\!\end{array}\right)_R - \frac{1}{2} X .
\end{equation}
The spontaneous breaking to the SM gauge group times $P_6$ can
be achieved by turning on a vacuum expectation value (VEV) in the upper $\SU2_R$ component of a 
$(\boldsymbol{4},\boldsymbol{1},\boldsymbol{2})_7$ and the lower $\SU2_R$ 
component of a $(\overline{\boldsymbol{4}},\boldsymbol{1},\boldsymbol{2})_{-7}$. 
With respect to \SU4 the VEV points always into the fourth direction 
such that the colour \SU3 remains unbroken. Since these components carry 
$\U1_{P_6}$ charges $\pm 6$ (cf. eqn.~(\ref{eqn:P6inPS})), their VEVs leave 
$P_6$ unbroken. However, in a supersymmetric theory, we encounter mixed 
$\SU2_{L/R}^2 \U1_X$ anomalies. For three families, these can be 
cancelled e.g.\ with the additional multiplets
\begin{equation}\label{eq:anomcan}
2\times(\boldsymbol{1},\boldsymbol{2},\boldsymbol{1})_{-6}, \quad 2\times(\boldsymbol{1},\boldsymbol{1},\boldsymbol{2})_{6}\; .
\end{equation}
These can have $\SU3 \times \SU2_L \times \U1_Y \times P_6$
invariant mass terms and hence decouple at the breaking scale. 

\subsection{Proton Hexality from $\boldsymbol{\text{SO}(12)}$}

As far as gauge coupling unification is concerned Pati-Salam is still
characterised by three couplings and, in that sense, not quite a GUT
yet. If we try to go one step further to \SO{10}, for instance, we are
back at the problems discussed in \cite{Luhn:2007gq}: the matter
fields in eqn.~(\ref{eq:assign}) should merge into a $\boldsymbol{16}$ of
\SO{10}. However, they cannot do that due to their opposite $\U1_X$
charges. The only way out, is to double the number of 16-dimensional
representations. Then, obviously, only half of each representation gives
rise to SM matter. However, for the remaining half representation 
not needed for matter a mechanism of {\it multiplet splitting} has to be invoked. 

With that in mind we might as well consider gauge groups larger than
\SO{10}  also accommodating the extra $\U1_X$. One canonical choice
is \SO{12}. For this discussion it is more convenient to use
Cartan--Weyl notation. There, a Lie algebra is given in the form
\begin{equation}
\left[ H_i, H_j\right] = 0 \;\;\; ,\;\;\; \left[ H_i ,
E_p\right] = p_i E_p ,
\end{equation}
with $i = 1,\ldots \mbox{rank}(G)=6$ and the $p$'s denote charge
vectors, or roots, of the remaining generators. \SO{12} has six
Cartan generators $H_1,\ldots, H_6$ and the roots $p$ are given by
\begin{equation}
 \left( \underline{\pm 1, \pm 1, 0,0,0,0}\right) .
\end{equation}
Here, all 60 roots are generated by permuting underlined entries, resulting 
in the $60+6=66$ dimensional adjoint of \SO{12}. Apart from the adjoint, the
following \SO{12} representations, specified by their weights, 
will be of interest for us
\begin{equation}\label{eq:so12reps}
\begin{array}{l l l}
\mbox{$\boldsymbol{32}$:} & \left( \pm \frac{1}{2}, \pm \frac{1}{2}, \pm
  \frac{1}{2}, \pm \frac{1}{2}, \pm \frac{1}{2}, \pm
  \frac{1}{2}\right) & \mbox{(even number of $-$ signs)},\\
\mbox{$\boldsymbol{32}^\prime$:} & \left( \pm \frac{1}{2}, \pm \frac{1}{2}, \pm
  \frac{1}{2}, \pm \frac{1}{2}, \pm \frac{1}{2}, \pm
  \frac{1}{2}\right) & \mbox{(odd number of $-$ signs)},\\
\mbox{$\boldsymbol{12}$:} & \left( \underline{\pm 1, 0,0,0,0,0}\right) .
& \end{array} 
\end{equation}   
The $\SU4\times \SU2_L\times \SU2_R \times \U1_X$ gauge group can
be embedded into the adjoint of \SO{12} as follows
\begin{equation}
\label{eqn:PSinSO12}
\begin{array}{l l}
\text{SU}(4):\,\,\, \left(\underline{\pm 1, \pm 1 , 0},0,0,0\right) , H_1,
H_2, H_3, &  \SU2_L:\,\,\, \pm \left( 0,0,0, 1, -1,0\right), H_4 -
H_5 ,\\
\SU2_R:\,\,\, \pm\left( 0,0,0,1,1,0\right) , H_4 + H_5 , & \text{U}(1)_X:\,\,\,
-2H_6 .
\end{array}
\end{equation}
Quarks and leptons become part of $\boldsymbol{32}$ and $\boldsymbol{32}^\prime$
  representations,
\begin{equation}
\begin{array}{l l l}
\mbox{$(\boldsymbol{4},\boldsymbol{2},\boldsymbol{1})_1$} : & (\underbrace{\textstyle \pm
  \frac{1}{2},\pm \frac{1}{2},\pm 
  \frac{1}{2}}_{\mbox{\scriptsize even \# $-$}},\underbrace{\textstyle
  \pm \frac{1}{2},\pm
  \frac{1}{2}}_{\mbox{\scriptsize odd \# $-$}},- \frac{1}{2})
& \subset \boldsymbol{32} \\
\mbox{$(\overline{\boldsymbol{4}},\boldsymbol{1},\boldsymbol{2})_{-1}$} : & (\underbrace{\textstyle \pm
  \frac{1}{2},\pm \frac{1}{2},\pm 
  \frac{1}{2}}_{\mbox{\scriptsize odd \# $-$}},\underbrace{\textstyle
  \pm \frac{1}{2},\pm 
  \frac{1}{2}}_{\mbox{\scriptsize even \# $-$}}, \frac{1}{2})
& \subset \boldsymbol{32}^\prime\;, \end{array}
\end{equation}
for each generation. Last, the electroweak Higgs sits inside a 12-dimensional representation
\begin{equation}
(\boldsymbol{1},\boldsymbol{2},\boldsymbol{2})_0: \quad ( 0,0,0,\underline{\pm 1 , 0}, 0)
\subset \boldsymbol{12}\;.
\end{equation}
Clearly, these multiplets have a fair amount of additional fields
which need to decouple via {\it multiplet splitting}. The situation
becomes much more dramatic for the remaining fields:
$(\boldsymbol{4},\boldsymbol{1},\boldsymbol{2})_7$, 
$(\overline{\boldsymbol{4}},\boldsymbol{1},\boldsymbol{2})_{-7}$ and 
the fields in eqn.~(\ref{eq:anomcan}). These are embedded in $\text{SO}(12)$
representations containing weights of the form $\left(\pm7/2,\ldots, \pm
  7/2\right)$ and $\left(0,\ldots, 0,\pm3\right)$, respectively. The big
charges under one of the $\text{SO}(12)$ Cartan generators can be
accommodated only in rather high-dimensional
representations.\footnote{The corresponding highest 
  weights are $\left( \frac{7}{2}, \frac{7}{2}, \frac{7}{2},
    \frac{7}{2}, \frac{7}{2}, \frac{7}{2}\right)$ and $\left(
    3,0,0,0,0,0\right)$ with Dynkin labels $\left[ 0,0,0,0,0,7\right]$
  and $\left[ 3,0,0,0,0,0\right]$. The dimensions can be computed
  using the Weyl formula (see e.g.~\cite{Slansky:1981yr}) and
  are 2,617,472 and 352. 
} Even with a working mechanism for
multiplet splitting one would rather not add such representations to
an underlying fundamental theory.{\footnote{Note that in the case of matter
parity as a subgroup of SO(10) we would need a 126-dimensional
representation to get the desired symmetry breakdown.
}}
Therefore, we should also have the
option of adding incomplete or {\it split multiplets}. In the next
section we will demonstrate that both mechanisms can be naturally obtained
within string theory.

\section{Proton Hexality and Local Grand Unification}  
\label{sec:ProHexfromlocalGUT}

The picture of local grand unification has drawn considerable
attention in the recent past in particular in the context of heterotic
model building (for recent reviews, see~\cite{Nilles:2008gq,Nilles:2009yd}). Before
discussing orbifold compactifications of the heterotic string, let us
demonstrate how the previous problems are solved in a bottom-up approach
with two extra dimensions\footnote{In the spirit of \cite{Forste:2004ie}
 this may correspond to a compactification of the heterotic string on
 $T^6/{\mathbb Z}_M\times \mathbb{Z}_N$, where our $T^2$ appears as a
fixed torus under the ${\mathbb Z}_N$ factor. For a list of possible
${\mathbb Z}_M \times {\mathbb Z}_N$ orbifolds see e.g.\
\cite{Font:1988mk}.}. For simplicity, we construct the
previously described Pati--Salam times $\U1_X$ model in four
dimensions. When discussing actual string models, later on, we
will be interested in four-dimensional gauge symmetry of the form:
SM gauge group times $P_6$ (times hidden sector gauge
group). Since the step from Pati--Salam times $\U1_X$ to the SM times $P_6$ 
is  comparatively simple there is no conceptual
difference.  

\begin{figure}[h]
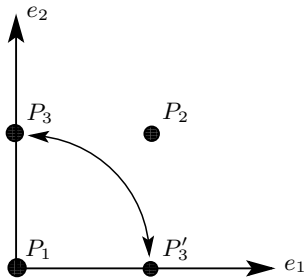

\begin{center}
\input fix.pstex_t
\end{center}
\caption{Fixed point structure of $T^2/{\mathbb Z}_4$ orbifold. 
The origin $P_1$  is fixed under any rotation. $P_2$ is fixed under
a $\pi/2$ rotation followed by a shift by $e_1$. $P_3$ and
$P_3^\prime$ are fixed under
a $\pi$ rotation and shifts by $e_2$  and $e_1$, respectively. $P_3$
and $P_3^\prime$ are related by a $\pi /2$  rotation and thus
identical on the orbifold.\label{fig:fix}}
\end{figure}

\begin{figure}[h]
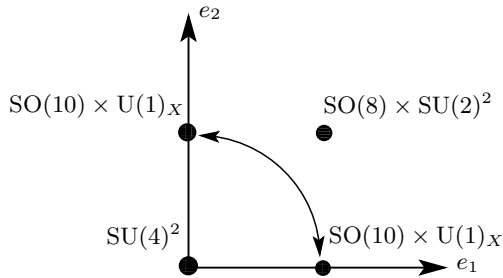

\begin{center}
\input geo.pstex_t
\end{center}
\caption{Gauge group geography of $T^2/{\mathbb Z}_4$ orbifold. The
gauge group in the bulk is \SO{12}.\label{fig:geo}}
\end{figure}

Two extra dimensions are compactified on a torus obtained by modding
the complex plane with a quadratic lattice spanned by the vectors 
$e_1$ and $e_2$. Further we mod out a ${\mathbb Z}_4$ symmetry generated 
by $\pi/2$ rotations in the plane. The fixed point structure is depicted 
in Figure~\ref{fig:fix}. 
As a six-dimensional theory we take an \SO{12} gauge theory. The
$\pi/2$ rotation is embedded into \SO{12} by the adjoint action of
$e^{2\pi\mbox{\scriptsize i}H_j V_j}$ and lattice shifts by $e_1$ or $e_2$ by
  $e^{2\pi\mbox{\scriptsize i}H_j W_j}$. (Shifts by $e_1$ and $e_2$ have
    to be embedded identically since $e_1$ is mapped onto $e_2$ by the
    $\pi/2$ rotation.) We choose
\begin{equation}
V= \left( \frac{1}{2}, \frac{1}{2},\frac{1}{2},0,0,0\right) \,\,\,
,\,\,\,
W = \left( 0,0,0,0,0,\frac{1}{2}\right) .
\end{equation}
The unbroken gauge group in four dimensions is the subgroup of
\SO{12} which is invariant under the orbifold action and
lattice shifts, i.e. $p \cdot V = 0 \text{ mod } 1$ and $p \cdot W = 0
\text{ mod } 1$, $p$ being a root of \SO{12}.  This yields
Pati-Salam times $\U1_X$ embedded into \SO{12} as
discussed in the previous section eqn.~(\ref{eqn:PSinSO12}). To avoid
enormous representations, the
$(\boldsymbol{4},\boldsymbol{1},\boldsymbol{2})_7$ and
$(\overline{\boldsymbol{4}},\boldsymbol{1},\boldsymbol{2})_{-7}$ and
fields in (\ref{eq:anomcan}) should be localised to points where
$\U1_X$ factorises and a big charge does not need a large
representation. The gauge group geography showing the local
projections of \SO{12} at the fixed points is depicted in
Figure \ref{fig:geo}.
At the fixed point(s) with $\SO{10} \times \U1_X$ gauge
group we localise a $2\times\boldsymbol{10}_{6}$, $2\times\boldsymbol{10}_{-6}$, {\bf 16}$_7$,
{\boldmath $\overline{16}$}$_{-7}$. From an \SO{12}
perspective these are {\it split multiplets} of dramatically lower
dimension than complete multiplets containing fields of the same
$\U1_X$ charges.  We demonstrate the mechanism of {\it
  multiplet splitting} by getting SM matter and Higgses
from complete \SO{12} multiplets in the bulk.  For each family
of quarks and leptons we add a $\boldsymbol{32}$ and a 
$\boldsymbol{32}^\prime$-plet and for the electroweak Higgs a $\boldsymbol{12}$-plet 
to the bulk. The action of orbifold and lattice shifts is again given by
$e^{2\pi\mbox{\scriptsize i}H_j V_j}$ and $e^{2\pi\mbox{\scriptsize
    i}H_j W_j}$ in the corresponding representation. In addition it is
accompanied by phase factors called orbifold parities. (For a
discussion within the context of heterotic orbifolds, see
e.g.~\cite{Kobayashi:2004ud}.) In our bottom-up approach we pick the
phases by hand. With an appropriate choice, it is easy to project the
multiplets exactly to the desired $\SU4 \times
\SU2_L\times \SU2_R \times \U1_X$ multiplets
in the four-dimensional effective theory.  There are still six-dimensional 
bulk anomalies to worry about. The relevant formul\ae\ can
be found e.g.\ in \cite{Erler:1993zy}. To cancel the bulk anomalies, we
have to add 16 $\boldsymbol{12}$-plets to the bulk theory if
all three families originate from bulk multiplets. We will not get
further into the details of our illustrative example. After all, this
theory is not UV complete and what we are really after is a
fully-fledged string model.

\section{Embedding Proton Hexality in Heterotic Orbifolds}  

\subsection{Proton Hexality from local GUTs}

In this section, we are interested in how to obtain string compactifications
furnished with gauged proton hexality.
We follow the strategy depicted in the previous section, i.e. we search for
MSSM-like constructions in which $P_6$ arises as a subgroup of
$\SU4\times\SU2_L\times\SU2_R\times\U1_X \subset \SO{12}$ and the matter generations
reside in (split) representations $\boldsymbol{32}$ and $\boldsymbol{32'}$
of the six-dimensional \SO{12} GUT.
With this purpose, we consider the $T^6/\mathbb{Z}_4 \times \mathbb{Z}_4$ orbifold
with torus lattice $\SO5^3$. The $\mathbb{Z}_4 \times \mathbb{Z}_4$ action is generated by the
twists
\begin{equation}
\label{eqn:Z4xZ4twists}
v_1 = \big(\tfrac{1}{4},0,-\tfrac{1}{4}\big) \quad\text{and}\quad v_2 = \big(0,\tfrac{1}{4},-\tfrac{1}{4}\big)\;. 
\end{equation}
This orbifold allows for one Wilson line of order 2 per \SO5 torus, hence three independent Wilson lines. 
Note that, since the twist $v_1$ ($v_2$) leaves the second (first) \SO5 torus invariant, it defines 
a $T^4/\mathbb{Z}_4$ orbifold.

Along the lines of the heterotic Mini-Landscape~\cite{Lebedev:2006kn,Lebedev:2007hv},
we apply a search strategy based on local GUTs to find a 
fertile region of heterotic orbifolds endowed with proton hexality.
The search strategy reads as follows. To start with, we consider the first \Z4. 
From all its possible gauge embeddings $V_1$~\cite{Stieberger:1998yi,Honecker:2006qz},
we choose the shift vector (denoted by (IVg) in~\cite{Honecker:2006qz})
\begin{equation}
\label{eqn:SO12shiftV1}
V_1 = \big( \tfrac{3}{4}, \tfrac{1}{4},0^6\big)\big(0^8\big)\;,
\end{equation}
which results in a $T^4/\mathbb{Z}_4$ model with \SO{12} gauge group
and twisted matter states transforming as $\boldsymbol{32}$ 
or $\boldsymbol{32'}$-plets. 
We look for all admissible gauge embeddings $V_2$ associated to the second 
$\mathbb{Z}_4$ action, resulting in 164 possibilities and select only those that break \SO{12} to Pati-Salam,
\begin{equation}
\SO{12} \quad \substack{V_2 \\ \longrightarrow} \quad \SU4 \times \SU2_L \times \SU2_R \times \U1_X\;.
\end{equation}
This reduces the number of promising shifts $V_2$ dramatically from 164 to 30. In
addition, we want SM matter representations arising from
$\boldsymbol{32}$ and $\boldsymbol{32'}$-plets, i.e. quarks and
leptons shall stem from the twisted sectors of the underlying
$T^4/\mathbb{Z}_4$ orbifold. Hence, we choose only those 12 shifts $V_2$ out of the 
30 where the $(i,0)$-twisted sector ($i = 1,2,3$) of the full
$T^6/\mathbb{Z}_4 \times \mathbb{Z}_4$ contains some Pati-Salam matter
representations.

The final step is to select models with up to two Wilson lines that 
yield the gauge group and matter spectrum of the MSSM plus 
vector-like exotics. With the help of the methods developed in~\cite{Lebedev:2008un},
we find more than 850 heterotic orbifolds with the desired properties. In many of these models, there is a \U1 that
by construction, leads to (at least some) quarks and leptons with the correct $P_6$ charges
and, therefore, suppression of proton decay occurs automatically.
However, the models we were able to construct suffer from exotics that are not vector-like w.r.t. 
$P_6$. In what follows we present an example of these constructions.

\paragraph{An example}

The model is defined by the shift $V_1$ of eqn.~\eqref{eqn:SO12shiftV1}, the second shift
\begin{equation}
V_2 = \big(-\tfrac{3}{4}, \tfrac{1}{2},-\tfrac{1}{2}, 0^3, \tfrac{1}{4}, \tfrac{1}{2}\big)\big(-1, 0^3, \tfrac{1}{4}^4\big)\;,
\end{equation}
and one Wilson line $W_2$ associated to the $e_2$ direction
\begin{equation}
W_2 = \big(-\tfrac{3}{4}, \tfrac{9}{4},-\tfrac{5}{4},-\tfrac{3}{4},-\tfrac{1}{4},\tfrac{5}{4},\tfrac{1}{4},-\tfrac{3}{4}\big)\big(\tfrac{1}{2}^4, 0^3, 2\big)\;.
\end{equation}
Schematically, the gauge symmetry is broken as expected:
\begin{equation}
\E8 \stackrel{V_1}{\longrightarrow} \SO{12} \stackrel{V_2}{\longrightarrow} \text{PS}\times\U1_X \stackrel{W_2}{\longrightarrow} \text{SM} 
\end{equation}
resulting in the four-dimensional gauge group
\begin{equation}
\SU3\times\SU2\times\U1_Y\times\U1^4\times[\SU4\times\SO{10}]\;.
\end{equation}
One of the \U1's, orthogonal to $\U1_Y$, appears anomalous.

Using the $\U1_X$ direction orthogonal to Pati-Salam, but inside
\SO{12} we can define a (non-anomalous) generator along the
lines of eqn.~\eqref{eqn:P6inPS}. Explicitly, it reads
\begin{equation}
t_{P_6} = \big(0,0,2,-2,2,-2,2,-2 \big) \big( 0^8 \big)\;.
\end{equation}
This is our prime choice for a \U1 proton hexality. In fact, the
spectrum contains three generations of quarks and leptons with the
right charge assignments and four SM singlets with $\U1_{P_6}$
charge $6$ that can be used to break $\U1_{P_6}$ to its
discrete subgroup $P_6$,
\begin{equation}
  3 (\boldsymbol{3}, \boldsymbol{2})_{\tfrac{1}{6}, 0} + 3 (\overline{\boldsymbol{3}}, \boldsymbol{1} )_{-\tfrac{2}{3}, 1} + 3 (\overline{\boldsymbol{3}}, \boldsymbol{1} )_{\tfrac{1}{3}, -1} + 3 (\boldsymbol{1}, \boldsymbol{2} )_{-\tfrac{1}{2}, 4} + 3 (\boldsymbol{1}, \boldsymbol{1} )_{1, 1} + 3 (\boldsymbol{1}, \boldsymbol{1} )_{0, -3} + 4 (\boldsymbol{1}, \boldsymbol{1} )_{0, 6}\;.
\end{equation}
See Table~\ref{tab:spectrumz4Xz4} for the full massless matter 
spectrum. The spectrum contains in addition exotics that are vector-like with
respect to the SM. Unfortunately, these exotics turn out
to be chiral with respect to $P_6$. Hence,
they can only decouple once $P_6$ is broken. One would have to make sure
that this breakdown is compatible with the desired proton stability.
These questions will be studied in a future publication.

\subsection{Proton Hexality as Accidental Symmetry}

We have focused so far on exact gauge symmetries of orbifold
compactifications and how they can be broken to give rise to $P_6$.
Alternatively, one can consider the large set of approximate
\U1 symmetries that arise naturally in orbifold models and lead to
solutions of certain issues, such as the strong
CP-problem~\cite{Choi:2006qj,Choi:2009jt}.

The general procedure to address this question is as follows.
Considering the effective superpotential $\mathcal{W}$ truncated at a
given order, one has to identify the \U1 symmetries (other than the
gauged ones) under which $\mathcal{W}$ is invariant. Then one must
require that a linear combination $\U1_{P_6}$ of these \U1s provide
the correct charge assignments for all the SM-fields (see
Table~\ref{tab:Charges}) and that there be at least one SM-singlet
$\chi$ with charges $\pm 6$ whose VEV breaks $\U1_{P_6}$ down to $P_6$.
One must further enforce that the spectrum be three generations plus
vector-like matter also w.r.t. $\U1_{P_6}$.

We have performed a search of such symmetries among the models of the
$\mathbb{Z}_6$-II Mini-Landscape with two and three Wilson lines and
found no example displaying three generations with the standard $P_6$
charges.  If one relaxes this condition and requires correct $P_6$
only for the first two generations, there are some examples. However,
they do not satisfy all the requirements listed before. In the
following we present a model of this type. It is not clear whether
better models are possible in this scenario. In fact the strategy
employed in the construction of the Mini-Landscape~\cite{Lebedev:2006kn,Lebedev:2007hv} aiming 
preferentially at complete multiplets could be incompatible with
the incorporation of $P_6$ in a satisfactory way.

\paragraph{An example}

Let us consider the \Z6--II orbifold model described by the shift embedding
\begin{equation}
  \label{eq:VE61}
  V^{\E6,1} ~=~ \left(\tfrac{1}{6},\,-\tfrac{1}{3},\,-\tfrac{1}{2},\,0,\,0,\,0,\,0,\,0\right)\left(0,\,0,\,0,\,0,\,0,\,0,\,0,\,0\right)
\end{equation}
and the Wilson lines
\begin{subequations}
\label{eq:Wilsonlines}  
\begin{eqnarray}
  W_3  & = & \left(-\tfrac{5}{6},\,-\tfrac{7}{6},\,\tfrac{1}{2},\,\tfrac{1}{2},\,\tfrac{1}{2},\,-\tfrac{1}{2},\,-\tfrac{1}{2},\,-\tfrac{1}{2} \right) \left(0,\,0,\,\tfrac{1}{3},\,\tfrac{1}{3},\,\tfrac{1}{3},\,0,\,1,\,\tfrac23\right)\,,\\
  W_2 & = & \left( 1,\,\tfrac{1}{2},\,0,\,\tfrac12,\,\tfrac12,\,-\tfrac12,\,-1,\,0 \right)\left(-\tfrac{1}{4},\,\tfrac{3}{4},\,\tfrac{1}{4},\, \tfrac{1}{4},\, \tfrac{3}{4},\,-\tfrac{3}{4},\,-\tfrac{3}{4},\,\tfrac{3}{4}\right)\,,\\
  W_2' & = & \left(\tfrac{3}{4},\,\tfrac{3}{4},\,-\tfrac{1}{4},\, -\tfrac{1}{4},\, -\tfrac{1}{4},\,\tfrac{3}{4},\,\tfrac{1}{4},\,\tfrac{1}{4}\right)\left(-\tfrac{1}{4},\,-\tfrac{1}{4},\,-\tfrac{1}{4},\, -\tfrac{1}{4},\, -\tfrac{1}{4},\,\tfrac{1}{4},\,\tfrac{1}{4},\,\tfrac{3}{4}\right)\,.
\end{eqnarray}
\end{subequations}
The four-dimensional gauge group is $\SU3\times\SU2_L\times\U1_Y\times[\SU6\times\U1^7]$ with $\U1_Y\subset\SU5$, and the
massless spectrum includes three SM generations plus vector-like exotics with respect to the SM gauge group.

At order four in the superpotential, apart from the gauge \U1s, there appear 82 accidental \U1s. A linear
combination of all the available \U1s renders the following properties (cf. Table~\ref{tab:spectrumAccidental}):

 a) there is a set of SM singlets $\chi$ that break $\U1_{P_6}$ to $P_6$;

 b) two SM families (including right-handed neutrinos) have proper $P_6$ charges;

 c) a pair of Higgs fields $h_u$, $h_d$ have proper $P_6$ charges; and
 
 d) apart from a third generation with unwanted $P_6$ charges, all other exotic states are vector-like.

This model has the potential to forbid proton decay. However, 
there are still some questions to be analysed.
First, in this specific model,
operators of order five in the superpotential break explicitly $\U1_{P_6}$. 
Secondly, a large mass for the top-quark is forbidden. Finally, 
$\U1_{P_6}$ exhibits
anomalies such as 
$\tr Q_{P_6}^3 , \tr Q_{P_6}^2 Q_Y , \tr Q_{Y}^2 Q_{P_6} \neq0$, that would 
have to be cancelled. A search for a more realistic model is under way.
 

\section{Conclusions}

Proton decay is a crucial question in grand unified theories.
The proton should decay, but not too fast. In supersymmetric
theories we have to forbid dim-4 and -5 operators. Usual matter
parity is not enough as it still allows dangerous dim-5
operators. In this respect proton hexality is perfect. It forbids
all the couplings that we do not want and allows those we
need. 

For a long time, its incompatibility with grand unification was
thought to be a problem. The concept of ``local grand unification''
discussed in string theories (and theories of extra dimensions),
however, changes the picture. Split multiplets, solving already
the problem of doublet-triplet splitting and the question
of breakdown of the grand unified gauge group, come to rescue
and make hexality potentially compatible with models that have 
been constructed in the framework of (heterotic) string theory.

Still the search for a fully realistic model requires more
work and perhaps a dedicated search strategy. Models as discussed
e.g. in the Mini-Landscape~\cite{Lebedev:2006kn,Lebedev:2007hv} are not so well
suited here by construction, as they were based on the desire to 
have unified multiplets for the first two families. Hexality
would require a different approach and it would be desirable
to set up a general geometric picture that naturally incorporates
$P_6$.

So far we have learned some lessons from string theory. Hexality
can come from various sources. It could be
\begin{itemize}  
\item a subgroup of a non-anomalous symmetry,
\item a subgroup of an anomalous symmetry,
\item an accidental symmetry.
\end{itemize}
In fact, $P_6$ could just be an approximate symmetry that is valid
at the level of lower-dimensional operators or valid only for
part of the spectrum (like the first and second family).

It is worthwhile to explore these questions further, both from
the bottom-up and top-down approaches. The present model building
attempts just scratch the surface of the landscape. More dedicated
model building is needed to construct fully realistic models.
We have here presented some toy models that illustrate the 
potential marriage of hexality with local grand unification. We are 
confident to report about the construction of more realistic models 
in the near future.

\section*{Acknowledgments}

P.V. would like to thank LMU Excellent for support.
S.R-S. is grateful to P.~Frampton, K.~Wang and T.~Yanagida
for useful discussions at IPMU.
This research was supported by the DFG cluster of excellence
Origin and Structure of the Universe, 
the SFB-Tansregio TR33
"The Dark Universe" (Deutsche Froschungsgemeinschaft) and 
the European Union 7th network program "Unification in the 
LHC era" (PITN-GA-2009-237920).

\newpage
\appendix
\section{Spectra of Toy Models}
\label{app:tables}
\enlargethispage{0.9cm}
\vskip -0.7cm
\begin{table}[H]
\begin{center}
{\scriptsize
\begin{tabular}{|rlc|c|rlc|c|rlc|}
\cline{1-3}\cline{5-7}\cline{9-11}
 \# & Irrep &  && \# & Irrep & && \# & Irrep &  \\
\cline{1-3}\cline{5-7}\cline{9-11}
  4 & $( {\bf 3}, {\bf 2}, {\bf 1}, {\bf 1})_{(1/6,0)}$             & $q$        &&  1 & $( {\bf\overline{3}}, {\bf 1}, {\bf 1}, {\bf 1})_{( 1/3,-4)}$ & $\bar{d}'$ &&
  8 & $( {\bf 1}, {\bf 2}, {\bf 1}, {\bf 1})_{( 1/2, 2)}\phantom{1^{1^1}}$ & $\bar{\ell}'$\\
  1 & $( {\bf 3}, {\bf 2}, {\bf 1}, {\bf 1})_{( 1/6, 1)}$           & $q'$       &&  2 & $( {\bf 3}, {\bf 1}, {\bf 1}, {\bf 1})_{(-1/3, 2)}$           & $d'$       &&
  1 & $( {\bf 1}, {\bf 2}, {\bf 1}, {\bf 1})_{( 1/2, 4)}$ & $\bar{\ell}'$\\
  2 & $( {\bf\overline{3}}, {\bf 2}, {\bf 1}, {\bf 1})_{(-1/6, 2)}$ & $\bar{q}'$ &&  3 & $( {\bf 3}, {\bf 1}, {\bf 1}, {\bf 1})_{(-1/3,-2)}$           & $d'$       &&
  1 & $( {\bf 1}, {\bf 2}, {\bf 1}, {\bf 1})_{( 1/2,-2)}$ & $\bar{\ell}'$\\
\cline{1-3}\cline{9-11}
  3 & $( {\bf\overline{3}}, {\bf 1}, {\bf 1}, {\bf 1})_{(-2/3, 1)}$ & $\bar{u}$  &&  3 & $( {\bf 3}, {\bf 1}, {\bf 1}, {\bf 1})_{(-1/3,-1)}$           & $d'$       &&
  9 & $( {\bf 1}, {\bf 2}, {\bf 1}, {\bf 1})_{( 1/2,-1)}\phantom{1^{1^1}}$ & $h_u$ \\
\cline{1-3}\cline{5-7}
  3 & $( {\bf\overline{3}}, {\bf 1}, {\bf 1}, {\bf 1})_{( 1/3,-1)}$ & $\bar{d}$  &&  4 & $( {\bf 1}, {\bf 2}, {\bf 1}, {\bf 1})_{(-1/2, 4)}$           & $\ell$     &&
  9 & $( {\bf 1}, {\bf 2}, {\bf 1}, {\bf 1})_{(-1/2, 1)}\phantom{1^{1^1}}$ & $h_d$ \\
\cline{9-11}
  3 & $( {\bf\overline{3}}, {\bf 1}, {\bf 1}, {\bf 1})_{( 1/3, 1)}$ & $\bar{d}'$ &&  7 & $( {\bf 1}, {\bf 2}, {\bf 1}, {\bf 1})_{(-1/2,-2)}$           & $\ell$     &&
  3 & $( {\bf 1}, {\bf 1}, {\bf 1}, {\bf 1})_{(1, 1)}\phantom{1^{1^1}}$    & $\bar{e}$ \\
\cline{9-11}
  3 & $( {\bf\overline{3}}, {\bf 1}, {\bf 1}, {\bf 1})_{( 1/3,-2)}$ & $\bar{d}'$ &&  1 & $( {\bf 1}, {\bf 2}, {\bf 1}, {\bf 1})_{(-1/2, 2)}$           & $\ell'$    &&
 18 & $( {\bf 1}, {\bf 1}, {\bf 1}, {\bf 1})_{(0,-3)}\phantom{1^{1^1}}$    & $\bar{\nu}$ \\
  1 & $( {\bf\overline{3}}, {\bf 1}, {\bf 1}, {\bf 1})_{( 1/3, 2)}$ & $\bar{d}'$ &&  1 & $( {\bf 1}, {\bf 2}, {\bf 1}, {\bf 1})_{(-1/2,-1)}$           & $\ell'$    &&
 15 & $( {\bf 1}, {\bf 1}, {\bf 1}, {\bf 1})_{(0, 3)}$    &   \\
\cline{1-3}\cline{5-7}\cline{9-11}
\multicolumn{11}{c}{}\\
\cline{1-3}\cline{5-7}\cline{9-11}
  3 & $( {\bf 1}, {\bf 1}, {\bf 1}, {\bf 1})_{(0, 6)}$                & $\chi$    &&  4 & $( {\bf 1}, {\bf 1}, {\bf\overline{4}}, {\bf 1})_{(-1/2,-5/2)}$ &  $s^-$    &&
  1 & $( {\bf 1}, {\bf 1}, {\bf 1}, {\bf 10})_{(0, 2)}\phantom{1^{1^1}}$   & $y$\\
  1 & $( {\bf 1}, {\bf 1}, {\bf 1}, {\bf 1})_{(0,-6)}$                & $\chi$    &&  2 & $( {\bf 1}, {\bf 1}, {\bf 4}, {\bf 1})_{(1/2, 5/2)}$            &  $s^+$    &&
  2 & $( {\bf 1}, {\bf 1}, {\bf 1}, {\bf 10})_{(0,-2)}$   &  \\
 18 & $( {\bf 1}, {\bf 1}, {\bf 1}, {\bf 1})_{(0, 0)}$                & $s^0$     &&  4 & $( {\bf 1}, {\bf 1}, {\bf 4}, {\bf 1})_{(1/2,-7/2)}$            &  $s^+$    &&
  2 & $( {\bf 1}, {\bf 1}, {\bf 1}, {\bf 10})_{(0, 1)}$   &  \\
\cline{1-3}\cline{9-11}
  8 & $( {\bf 1}, {\bf 2}, {\bf\overline{4}}, {\bf 1})_{(0,-1/2)}$    & $m$       &&  4 & $( {\bf 1}, {\bf 1}, {\bf 4}, {\bf 1})_{(1/2,-1/2)}$            &  $s^+$    &&
  2 & $( {\bf 1}, {\bf 1}, {\bf 1}, {\bf 1})_{(0, 2)} \phantom{1^{1^1}}$    & $\tilde s$\\
  2 & $( {\bf 1}, {\bf 2}, {\bf 4}, {\bf 1})_{(0,-5/2)}$              & $m'$      &&  2 & $( {\bf 1}, {\bf 1}, {\bf\overline{4}}, {\bf 1})_{(1/2, 1/2)}$  &  $s^+$    &&
  7 & $( {\bf 1}, {\bf 1}, {\bf 1}, {\bf 1})_{(0,-2)}$    &  \\
\cline{1-3}\cline{5-7}
  6 & $( {\bf 1}, {\bf 1}, {\bf 4}, {\bf 1})_{(-1/2, 3/2)}$           &  $s^-$    &&  5 & $( {\bf 1}, {\bf 1}, {\bf 6}, {\bf 1})_{(0, 2)}$                & $x$ &&
  3 & $( {\bf 1}, {\bf 1}, {\bf 1}, {\bf 1})_{(0, 1)}\phantom{1^{1^1}}$    &  \\
  2 & $( {\bf 1}, {\bf 1}, {\bf 4}, {\bf 1})_{(-1/2,-3/2)}$           &  $s^-$    &&  6 & $( {\bf 1}, {\bf 1}, {\bf 6}, {\bf 1})_{(0, 1)}$                &     &&
  9 & $( {\bf 1}, {\bf 1}, {\bf 1}, {\bf 1})_{(0,-1)}$    &  \\
\cline{1-3}\cline{5-7}\cline{9-11}
\end{tabular}
\caption{Massless spectrum of a model with gauged $P_6$. Quantum numbers w.r.t.
  $[\SU3_C\times\SU2_\mathrm{L}]\times[\SU4\times\SO{10}]$ (bold) and $\U1_Y\times\U1_{P_6}$ (subscripts) are given.}
\label{tab:spectrumz4Xz4}
}
\end{center}
\end{table}
\vskip -1.2cm
\begin{table}[H]
\begin{center}
{\scriptsize
\begin{tabular}{|rlc|rlc|c|rlc|}
\cline{1-6}\cline{8-10}
  \#  &  Irrep                                          &  & \# & Anti-irrep       &     &&  \#  & Irrep &  \\
\cline{1-6}\cline{8-10}
  2 & $( {\bf 3}, {\bf 2};  {\bf 1})_{(1/6, 0)}$   & $q_{1,2}$ &
    & $\phantom{1^{1^1}}$ & &&
 11 & $( {\bf 1}, {\bf 1};  {\bf 1})_{(0, 0)}$     & $s^0$ \\
  1 & $( {\bf 3}, {\bf 2};  {\bf 1})_{(1/6, 25/2)}$   & $q_{3}$ &
    & & &&
  4 & $( {\bf 1}, {\bf 1};  {\bf 1})_{(0, \pm6)}$     & $\chi$ \\
  1 & $( {\bf 3}, {\bf 2};  {\bf 1})_{(1/6, 7/2)}$   & $q'$ &
  1 & $( {\bf\overline{3}}, {\bf 2};  {\bf 1})_{(-1/6, -7/2)}$  & $\bar q'$ &&
  6 & $( {\bf 1}, {\bf 1};  {\bf 1})_{(0, 3)}$     &  $\bar{\nu}$  \\
\cline{1-6}\cline{8-10}
  2 & $( {\bf 1}, {\bf 2};  {\bf 1})_{(-1/2, -2)}$  & $\ell_{1,2}$ &
    & & &&
  4 & $( {\bf 1}, {\bf 1};  {\bf 1})_{(0, \pm21/31)}$     &   $\tilde{s}$  \\
  1 & $( {\bf 1}, {\bf 2};  {\bf 1})_{(-1/2, 0)}$  & $\ell_{3}$ &
    & & &&
  4 & $( {\bf 1}, {\bf 1};  {\bf 1})_{(0, \pm1/2)}$     &   \\
  1 & $( {\bf 1}, {\bf 2};  {\bf 1})_{(-1/2, 1)}$  & $h_d$ &
  1 & $( {\bf 1}, {\bf 2};  {\bf 1})_{(1/2, -1)}$  & $h_u$ &&
  4 & $( {\bf 1}, {\bf 1};  {\bf 1})_{(0, \pm3/2)}$     &   \\
  2 & $( {\bf 1}, {\bf 2};  {\bf 1})_{(-1/2, 2)}$  & $\ell'$ &
  2 & $( {\bf 1}, {\bf 2};  {\bf 1})_{(1/2, -2)}$  & $\bar\ell'$ &&
  2 & $( {\bf 1}, {\bf 1};  {\bf 1})_{(0, \pm7/2)}$     &   \\
  1 & $( {\bf 1}, {\bf 2};  {\bf 1})_{(-1/2, -1/2)}$  &  &
  1 & $( {\bf 1}, {\bf 2};  {\bf 1})_{(1/2, 1/2)}$  &  &&
  2 & $( {\bf 1}, {\bf 1};  {\bf 1})_{(0, \pm2/9)}$     &   \\
  1 & $( {\bf 1}, {\bf 2};  {\bf 1})_{(-1/2, 3)}$  &  &
  1 & $( {\bf 1}, {\bf 2};  {\bf 1})_{(1/2, 3)}$  &  &&
  2 & $( {\bf 1}, {\bf 1};  {\bf 1})_{(0, \pm7/9)}$     &   \\
\cline{1-6}
  2 & $( {\bf\overline3}, {\bf 1};  {\bf 1})_{(1/3, -1)}$  & $\bar d_{1,2}$ &
    & & &&
  2 & $( {\bf 1}, {\bf 1};  {\bf 1})_{(0, \pm52/27)}$     &   \\
  1 & $( {\bf\overline3}, {\bf 1};  {\bf 1})_{(1/3, 0)}$  & $\bar d_{3}$ &
    & & &&
  2 & $( {\bf 1}, {\bf 1};  {\bf 1})_{(0, \pm31/27)}$     &   \\
  3 & $( {\bf\overline3}, {\bf 1};  {\bf 1})_{(1/3, 2)}$  & $\bar d'$ &
  3 & $( {\bf3}, {\bf 1};  {\bf 1})_{(-1/3, -2)}$  & $d'$ &&
  2 & $( {\bf 1}, {\bf 1};  {\bf 1})_{(0, \pm22/27)}$     &   \\
  1 & $( {\bf\overline3}, {\bf 1};  {\bf 1})_{(1/3, 0)}$  &  &
  1 & $( {\bf3}, {\bf 1};  {\bf 1})_{(-1/3, 6)}$  &  &&
  2 & $( {\bf 1}, {\bf 1};  {\bf 1})_{(0, \pm14/27)}$     &   \\
  1 & $( {\bf\overline3}, {\bf 1};  {\bf 1})_{(1/3, 1/2)}$  &  &
  1 & $( {\bf3}, {\bf 1};  {\bf 1})_{(-1/3, -1/2)}$  &  &&
  2 & $( {\bf 1}, {\bf 1};  {\bf 1})_{(0, \pm2/27)}$     &   \\
\cline{1-6}
  2 & $( {\bf\overline3}, {\bf 1};  {\bf 1})_{(-2/3, 1)}$  & $\bar u_{1,2}$ &
    & & &&
  2 & $( {\bf 1}, {\bf 1};  {\bf 1})_{(0,\pm125/27 )}$     &   \\
  1 & $( {\bf\overline3}, {\bf 1};  {\bf 1})_{(-2/3, 9/2)}$  & $\bar u_{3}$ &
    & & &&
  2 & $( {\bf 1}, {\bf 1};  {\bf 1})_{(0, \pm244/27)}$     &   \\
  1 & $( {\bf\overline3}, {\bf 1};  {\bf 1})_{(-2/3, -7/9)}$  & $\bar u'$ &
  1 & $( {\bf3}, {\bf 1};  {\bf 1})_{(2/3, 7/9)}$  & $u'$ &&
  2 & $( {\bf 1}, {\bf 1};  {\bf 1})_{(0, \pm8/31)}$     &   \\
\cline{1-6}
  3 & $( {\bf1}, {\bf 1};  {\bf 1})_{(1, 1)}$  & $\bar e_{1,2,3}$ &
    & & &&
  2 & $( {\bf 1}, {\bf 1};  {\bf 1})_{(0, \pm1/31)}$     &   \\
  1 & $( {\bf1}, {\bf 1};  {\bf 1})_{(1, 0)}$  & $\bar e'$ &
  1 & $( {\bf1}, {\bf 1};  {\bf 1})_{(-1, 12)}$  & $e'$ &&
  2 & $( {\bf 1}, {\bf 1};  {\bf 1})_{(0, \pm16/31)}$     &   \\
\cline{1-6}
  1 & $( {\bf 3},  {\bf 1};  {\bf 1})_{(1/6, 25/31)}$   & $v$ &
  1 & $( {\bf\overline{3}}, {\bf 1};  {\bf 1})_{(-1/6, -25/31)}$  & $\bar v$ &&
  1 & $( {\bf 1}, {\bf 1};  {\bf 1})_{(0, -34/9)}$     &   \\
\cline{1-6}
  1 & $( {\bf 1},  {\bf 1};  {\bf\overline 6})_{(1/2, 40/27)}$   & $w^+$ &
  1 & $( {\bf 1}, {\bf 1};   {\bf 6})_{(-1/2, -40/27)}$ &  $w^-$ &&
  1 & $( {\bf 1}, {\bf 1};  {\bf 1})_{(0, -20/9)}$     &   \\
\cline{1-6}
  3 & $( {\bf 1},  {\bf 1};  {\bf 1})_{(1/2, 2/27)}$   & $s^+$ &
  3 & $( {\bf 1},  {\bf 1};  {\bf 1})_{(-1/2, -2/27)}$  & $s^-$ &&
  1 & $( {\bf 1}, {\bf 1};  {\bf 1})_{(0, 29/3)}$     &   \\
  2 & $( {\bf 1},  {\bf 1};  {\bf 1})_{(1/2, -2)}$   & &
  2 & $( {\bf 1},  {\bf 1};  {\bf 1})_{(-1/2, 8)}$  &  &&
  1 & $( {\bf 1}, {\bf 1};  {\bf 1})_{(0, -11/3)}$     &   \\
\cline{8-10}
  1 & $( {\bf 1},  {\bf 1};  {\bf 1})_{(1/2, 0)}$   &  &
  1 & $( {\bf 1},  {\bf 1};  {\bf 1})_{(-1/2, 0)}$  &  &&
  2 & $( {\bf 1}, {\bf 1};  {\bf6})_{(0, 0)}$    & $x$ \\
  1 & $( {\bf 1},  {\bf 1};  {\bf 1})_{(1/2, 1/31)}$   &  &
  1 & $( {\bf 1},  {\bf 1};  {\bf 1})_{(-1/2, -1/31)}$  &  &&
  2 & $( {\bf 1}, {\bf 1};  {\bf 6})_{(0, \pm1/2)}$    & $x'$ \\
\cline{8-10}
  1 & $( {\bf 1},  {\bf 1};  {\bf 1})_{(1/2, -1/2)}$   &  &
  1 & $( {\bf 1},  {\bf 1};  {\bf 1})_{(-1/2, 1/2)}$  &  &&
  2 & $( {\bf 1}, {\bf 1};  {\bf\overline{6}})_{(0, 0)}\phantom{1^{1^1}}$    & $\bar{x}$\\
  1 & $( {\bf 1},  {\bf 1};  {\bf 1})_{(1/2, 2)}$   &  &
  1 & $( {\bf 1},  {\bf 1};  {\bf 1})_{(-1/2, -2)}$  &  &&
  2 & $( {\bf 1}, {\bf 1};  {\bf\overline{6}})_{(0, \pm1/2)}$  & $\bar{x}'$ \\
\cline{1-6}\cline{8-10}
  2 & $( {\bf 1},  {\bf 2};  {\bf 1})_{(0, 0)}$     & $m$ &  \multicolumn{3}{c|}{$\phantom{I^{I^I}}$} &  \multicolumn{3}{c}{$\phantom{I^{I^I}}$} \\
  2 & $( {\bf 1},  {\bf 2};  {\bf 1})_{(0, 3/31)}$  & $m'$ & 2 & $( {\bf 1},  {\bf 2};  {\bf 1})_{(0, -3/31)}$  & $\bar m'$  &  \multicolumn{3}{c}{$\phantom{I^{I^I}}$} \\
  1 & $( {\bf 1},  {\bf 2};  {\bf 1})_{(0, 1)}$  &  & 1 & $( {\bf 1},  {\bf 2};  {\bf 1})_{(0, -1)}$  &   &  \multicolumn{3}{c}{$\phantom{I^{I^I}}$} \\
  1 & $( {\bf 1},  {\bf 2};  {\bf 1})_{(0, -3)}$  & & 1 & $( {\bf 1},  {\bf 2};  {\bf 1})_{(0, 3)}$  &   &  \multicolumn{3}{c}{$\phantom{I^{I^I}}$} \\
\cline{1-6}
\end{tabular}
\caption{Massless spectrum of a model with $P_6$ as accidental symmetry. Quantum numbers w.r.t.
  $[\SU3_C\times\SU2_\mathrm{L}]\times[\SU6]$ (bold) and $\U1_Y\times\U1_{P_6}$ (subscripts) are given.}
\label{tab:spectrumAccidental}
}
\end{center}
\end{table}

\newpage
\providecommand{\bysame}{\leavevmode\hbox to3em{\hrulefill}\thinspace}

\end{document}